\documentclass[aps,prb,preprint,superscriptaddress]{revtex4}
\usepackage{graphicx}
\begin{document}

\title{Influence of the Rashba effect on
the Josephson current through a Superconductor/Luttinger
Liquid/Superconductor tunnel junction}
\author{I. V. Krive}
\affiliation{Department of Applied Physics, Chalmers University of
  Technology and G\"oteborg University, SE-412 96 G\"oteborg,
  Sweden}
  \affiliation{B. I. Verkin Institute for Low Temperature Physics and
Engineering, 47 Lenin Avenue, 61103 Kharkov, Ukraine}
\author{A. M. Kadigrobov}
\affiliation{Department of Applied Physics, Chalmers University of
  Technology and G\"oteborg University, SE-412 96 G\"oteborg,
  Sweden}
\affiliation{Theoretische Physik III, Ruhr-Universit\"{a}t Bochum,
D-44780 Bochum, Germany}
\author{R. I. Shekhter}
\affiliation{Department of Applied Physics, Chalmers University of
  Technology and G\"oteborg University, SE-412 96 G\"oteborg,
  Sweden}
\author{M. Jonson}
\affiliation{Department of Applied Physics, Chalmers University of
  Technology and G\"oteborg University, SE-412 96 G\"oteborg,
  Sweden}

\begin{abstract}
The Josephson current through a 1D quantum wire with Rashba
spin-orbit and electron-electron interactions is calculated. We
show that the interplay of Rashba and Zeeman interactions gives
rise to a supercurrent through the 1D conductor that is anomalous
in the sense that it persists in the absence of any phase
difference between the two superconducting leads to which it is
attached. The electron dispersion asymmetry induced by the Rashba
interaction in a Luttinger-liquid wire plays a significant role
for poorly transmitting junctions. It is shown that for a weak or
moderate electron-electron interaction the spectrum of plasmonic
modes confined to the normal part of the junction becomes
quasi-random in the presence of dispersion asymmetry.
\end{abstract}
%


\maketitle
\newpage

\section{Introduction}

In recent years the concept of a Luttinger liquid (LL) as a
realistic  model of interacting electrons in one-dimensional (1D)
metallic structures has received experimental support (see e.g.
Refs.~[\onlinecite{1,2,3,4,5}]). Quantum wires (QW) in laterally
constrained 2D electron gases (2DEG) and single wall carbon
nanotubes (SWNT) - are the two best known structures where LL
behavior has been established both theoretically and
experimentally.

In SWNTs, where the interaction effects have been shown to be
strong\cite{3,4,5}, interactions
 influence the charge and spin
transport through the nanotube. When a repulsively interacting LL
is coupled to leads (M) of noninteracting electrons, two
qualitatively different regimes of charge transport may be
realized depending on the quality of the LL/M electrical contacts.
For tunnel contacts charge transport through the system is
strongly suppressed at low temperatures and bias voltages\cite{KF}
by the repulsive electron-electron (e-e) interaction. In contrast,
for adiabatic contacts when electron backscattering is negligibly
small, the conductance is not renormalized by the
interaction\cite{7,8,9}.

These two types of charge transport behavior also characterize the
superconducting properties of a LL wire coupled to
superconductors. For adiabatic contacts only Andreev scattering of
electrons occurs at the boundaries between LL and bulk
superconductors (LL/S boundaries). This process does not lead to a
redistribution of charge density along the wire and therefore the
Coulomb interaction does not influence the supercurrent through a
perfect LL. The above result was proved in
Ref.~[\onlinecite{Maslov}] by a direct calculation of the
Josephson current through a long S/LL/S junction, $\;L\gg\xi_0\;,
(L$ is the junction length, $\xi_0=\hbar v_F/|\Delta|$ is the
superconducting coherence length, $\Delta$ is the superconducting
order parameter). In the case of tunnel S/I/LL/I/S junction ---
where "I" denotes the insulating "layer" --- the repulsive e-e
interaction results in a renormalization of the junction
transparency and the critical Josephson current is strongly
suppressed\cite{Fazio}.

Here we consider the influence of spin-orbit (s-o) interaction on
the Josephson current through a long S/I/LL/I/S junction. It has
been known for a long time that the s-o interaction is strong in a
2DEG formed in a GaAS/AlGaAs inversion layer (the Rashba effect
\cite{R}) and that it can be controlled by a gate
voltage\cite{13,14,15}. In what follows we will consider a quantum
wire in a laterally confined 2DEG coupled to superconducting
electrodes via tunnel barriers.

\begin{figure}
  \centerline{\includegraphics[width=8.0cm]{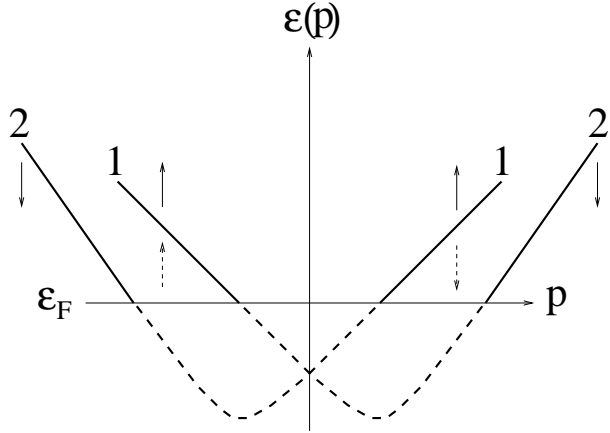}}
  \vspace*{2mm}
  \caption{Schematic energy spectrum of 1D spin-1/2 electrons with
   dispersion asymmetry. The subbands "1" and "2" are characterized
   by their Fermi velocities $v_{1F}\neq v_{2F}$. In the case of
   weak Rashba interaction spin projections in the subbands for
   each given momentum are opposite. For strong Rashba interaction
   spins in subbands are parallel and they are opposite for right-
   and left-moving particles.}
  \label{fig1}
  \end{figure}

The influence of the Rashba effect on the electron spectrum and on
the transport properties of quasi-1D quantum wires has been
studied theoretically in Refs.~[\onlinecite{M1,M2}], where it was
shown that the s-o interaction not only splits the electron
spectrum into spin-"up" and spin-"down" subbands, but additionally
breaks the chiral symmetry. This implies that left- and
right-moving electrons with the same spin projection have
different Fermi velocities. Since the time invariance (T-symmetry)
of the spin-orbit Hamiltonian implies that
$\,v_{R\uparrow}^{(F)}=v_{L\downarrow}^{(F)}=v_{1F}$ and
$\,v_{R\downarrow}^{(F)}=v_{L\uparrow}^{(F)}=v_{2F}$ the strength
of the Rashba effect in a single-channel QW can be characterized
by a dispersion asymmetry parameter
$\lambda_a=(v_{1F}-v_{2F})/(v_{1F}+v_{2F})$. In
Refs.~[\onlinecite{M1,M2}] it was assumed that in the presence of
Rashba interactions the electron spins in a quasi-1D wire are
aligned as in the 2D case (see Fig.\ref{fig1}, solid lines for spin
projections). Although this assumption is not valid for a strong
Rashba coupling\cite{GZ}, the model considered by the authors of
these references is interesting in itself. It allows one to study
the effects of dispersion asymmetry ($\lambda_a\neq 0$) on the
electron dynamics and in the limit $\lambda_a\rightarrow 0$ it
reproduces the standard results for spin-1/2 electrons without s-o
interaction.

Since the electron spin is not conserved in the presence of s-o
interactions the classification of spin states  assumed in
Refs.~[\onlinecite{M1,M2}] is not evidently correct. Actually, as
was shown in Ref.~[\onlinecite{GZ}], it can be justified only for
a weak Rashba interaction. In the most interesting case of strong
Rashba interaction, when the characteristic energy scale
introduced by s-o coupling is compared with the energy spacing of
1D subbands, the average spin projections for electrons with large
(Fermi) momentum are different. The total energy is minimized when
all right-moving (R) electrons have parallel spins and they are in
the opposite direction to the spins of left-moving (L)
electrons\cite{GZ}. In what follows we choose the sign of the
Rasha interaction so that R-electrons have "down"-spin and
L-electrons are "up"-spin particles (see Fig.\ref{fig1}, dashed lines for
spin projections). Notice that under conditions when the Rashba
effect is active the electron spin lies in a (2D) plane and
orthogonal to the electron momentum (in the 1D case this direction
is fixed and "up" and "down" spin projections are well defined).

\begin{figure}
  \centerline{\includegraphics[width=8.0cm]{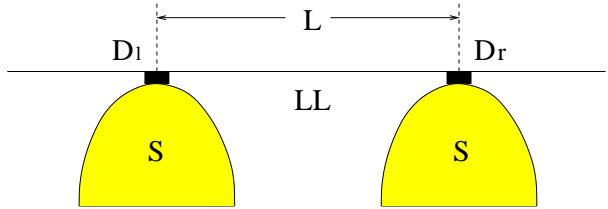}}
  \vspace*{2mm}
  \caption{SLLS-junction of lengh $L$ formed by an effectively
   infinite Luttinger liquid coupled to bulk superconductors by
   side electrodes.}
  \label{fig2}
  \end{figure}

At first we consider the influence of electron dispersion asymmetry
in the model elaborated in Refs.[\onlinecite{M1,M2}]
on the superconducting properties of S/I/LL/I/S junction.
We calculate Josephson current perturbatively on the junction
transparency $D=|t_{l}t_{r}|^2,\;(|t_{l,r}|^{2}\ll 1$ are the
transparencies of tunnel barriers at left and right LL/S
interfaces) and for arbitrary strength of electron-electron
interaction, dispersion asymmetry $\lambda_a$ and Zeeman
splitting $\Delta_Z=g\mu_B B\;(g$ is g-factor, $\mu_B$ is
Bohr magneton and $B$ is the magnetic field). Two different
geometries of S/LL/S junction are considered. In the first
case an effectively infinite LL is connected by the side
electrodes to the bulk superconductors (Fig.\ref{fig2}). In this
geometry\cite{Fazio} one can use periodic boundary conditions
for plasmons and
all calculations can be done analytically even in the presence
of s-o interaction. When dispersion asymmetry is negligibly
small ($\lambda_a\rightarrow 0$) we reproduce the formula for
Josephson current derived in Ref.[\onlinecite{Fazio}]. As was shown
in the cited paper the influence of Coulomb interaction on
a supercurrent through a tunnel junction $J=J_c\sin\varphi$
results in suppression of the critical current, which in the presence of
Zeeman splitting takes the form $J_c=J_{c}^{(0)}R_i(g_c)\cos(\Delta_Z/
\Delta_L)$, where $J_{c}^{(0)}\sim D\Delta_L\; (\Delta_L=\hbar v_F/L)$
is the critical current for noninteracting electrons and
the interaction-induced renormalization factor $R_i$ (the subindex
"i"  labels
the case of effectively {\em infinite} LL wire) is
small for repulsively interacting electrons $R_i(g_c\ll 1)
\ll 1$ (here $g_c$ is the LL correlation parameter in the
charge sector). The Zeeman interaction in the absence of Rashba
effect results only in additional sign alternating harmonic
factor in the critical current (see also Refs.[\onlinecite{19}]).

We assume that the magnetic field is local and influencing only
the normal (nonsuperconducting) part of the junction. (This can be
realized in an experiment for instance with the help of a magnetic
tip in a scanning tunnelling microscope). The interplay of
dispersion asymmetry and the Zeeman interaction leads to beatings
in the supercurrent considered as a function of the local magnetic
field $B$. A more unusual prediction is the appearance of
supercurrent $J_{a}$ even at $\varphi=0$. The existence of this
anomalous Josephson current is related to the breaking of chiral
invariance in quasi-1D quantum wires and the effect manifests
itself already for noninteracting particles (see
Ref.~[\onlinecite{KGSJ}]).

\begin{figure}
  \centerline{\includegraphics[width=8.0cm]{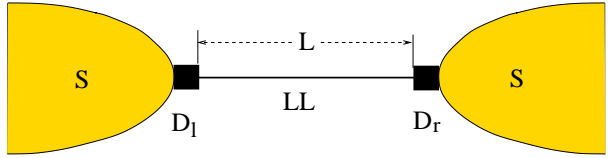}}
  \vspace*{2mm}
  \caption{A Luttinger liquid wire of length $L$ coupled to bulk
   superconductors via tunnel barriers with transparencies $D_{l(r)}$. }
  \label{fig3}
  \end{figure}

A more realistic geometry for an S/LL/S junction is a finite LL
wire (of the length $L$) coupled via tunnel barriers to bulk
superconductors (Fig.\ref{fig3}). We will assume that the barrier
transparencies are unequal and small (nonsymmetric tunnel junction
$|t_l|\ne |t_r|\ll 1$) and evaluate the $\varphi$-dependent part
of the ground state energy in perturbation theory using the
junction transparency $D=|t_{l}t_{r}|^2$ as expansion parameter.
Normal and Andreev scattering at the interfaces can be taken into
account by the boundary terms\cite{Zag} in the Hamiltonian of the
S/I/LL/I/S junction. To first order in the junction transparency
the problem is reduced to the evaluation of four-fermion
correlation functions for a two channel LL Hamiltonian with the
boundary conditions implying the absence of particle current
through the S/LL interfaces at $x=0,L$. In the absence of
spin-orbit interaction the problem of quantization of plasmon
modes in a finite LL with open ends was solved in
Ref.~[\onlinecite{FG}]. Here, we generalize the quantization
procedure proposed in the cited paper to the case of spin-1/2
fermions with dispersion asymmetry.

The two-channel LL Hamiltonian describing our system is
diagonalized exactly by the canonical transformation suggested in
Ref.~[\onlinecite{IVK}]. We show that the spectrum of plasmons in
a LL with open ends in the presence of dispersion asymmetry is
determined by a transcendental equation. In the general case the
spectrum forms a set of quasi-random energy levels. For
noninteracting electrons, or when the energy dispersions are
symmetric ($v_{1F}=v_{2F}=v_F$), the spectrum is reduced to a set
of equidistant energy levels. In the limit of strongly interacting
particles the plasmon spectrum also becomes regular. We calculate
the Josephson current in the cases when the spectral equation can
be solved analytically.

We find that for noninteracting electrons the critical Josephson
current through a tunnel S/QW/S junction is enhanced by the
presence of dispersion asymmetry. This behavior is specific for 1D
electrons and the effect disappears in 2D junctions\cite{BR}.

When the Rashba effect is not pronounced ($\lambda_a\rightarrow
0$), the Josephson current through a tunnel S/LL/S junction is
described by an expression analogous to the formula derived for an
effectively infinite LL. However, the interaction-induced
renormalization (suppression) of the critical supercurrent is much
stronger for a finite LL than for an infinite one $R_f(g_c\ll
1)\ll R_i(g_c\ll 1)$ provided the electron-electron interaction is
short-ranged.

As already mentioned in the Introduction, the electron spin states
in a 1D quantum wire in the regime of strong Rashba effect are
fully determined by s-o interaction and the electrons with large
(Fermi) momenta behave as truly chiral particles. That is the
electron spin polarizations and the direction of their motion
(right/left) are strongly correlated and all right(left)- moving
particles, irrespective of their Fermi velocities, have parallel
spins\cite{GZ} which are opposite to the spin polarizations of
left(right)-moving electrons. So it is reasonable to expect that
in this case the magnetic field via the Zeeman interaction will
induce an anomalous supercurrent (at $\varphi=0$) even in the
absence of dispersion asymmetry. We calculate the Josephson
current in a SILLIS junction in a model when the Rashba s-o
interaction is smoothly switched on in a 1D QW and spin-flips are
not accompanied by electron backscattering. An anomalous influence
of Zeeman splitting on the critical supercurrent is predicted.

\section{Proximity-induced superconductivity in
a Luttinger liquid wire with chiral symmetry breaking}

It is physically evident that the Coulomb interaction in a long
S/I/LL/I/S junction suppresses the critical supercurrent due to a
strong Kane-Fisher renormalization of the bare tunneling matrix
elements. The Josephson current through a Luttinger liquid coupled
to bulk superconductors via tunnel contacts was first calculated
by Fazio {\em et al.}\cite{Fazio} who showed that the critical
supercurrent is multiplicatively renormalized (suppressed) by a
repulsive electron-electron interaction. The calculations were
performed in linear (Fig.\ref{fig2}) and ring-like geometries. In both
cases periodic boundary conditions for the plasmonic modes can be
imposed. Although from an experimental point of view the
considered geometries of an SNS junction look rather artificial,
they do allow one to simplify the calculations.

For noninteracting electrons the critical supercurrents in an SNS
junction formed by a long (effectively infinite) quantum wire
connected to superconductors by side tunnel contacts (separated by
a distance $L$) and in an SNS junction where a finite length QW
bridges the gap (of the same length L) between two superconductors
differ only by a numerical factor. If the QW is treated as a
Luttinger liquid this factor becomes a function of the interaction
strength  and can be evaluated analytically (see below). When both
electron-electron interactions and dispersion asymmetry are
present the calculations are more cumbersome. We start with the
case of a side-contacted QW where we are able to analytically
evaluate the supercurrent for arbitrary interaction strength and
dispersion asymmetry parameter.

The Hamiltonian $H=H_{LL}+H_b$ of a S/I/LL/I/S junction is a sum
of the LL Hamiltonian $H_{LL}$ and the boundary Hamiltonian $H_b$.
The latter describes the effective boundary pairing and scattering
interactions produced by the superconducting and normal scattering
potentials at the points $x=0$ and $x=L$ (see
Ref.~[\onlinecite{Zag}]). When chiral symmetry is broken the
corresponding spin-1/2 LL Hamiltonian expressed in terms of charge
densities of chiral fields takes the form
\begin{eqnarray} \label{llh}
H_{LL}&=&\pi\hbar \int dx \{ u_{1}( \rho_{R\uparrow}^2+
\rho_{L\downarrow}^2)+
u_{2}(\rho_{R\downarrow}^2 +\rho_{L\uparrow}^2)\\
&+&\frac{V_{0}}{\pi\hbar}(\rho_{R\uparrow}\rho_{R\downarrow}+
\rho_{L\uparrow}\rho_{L\downarrow}
+\rho_{R\uparrow}\rho_{L\uparrow}+\rho_{R\downarrow}\rho_{L\downarrow}
\nonumber \\
&+&\rho_{R\uparrow}\rho_{L\downarrow}+\rho_{R\downarrow}\rho_{L\uparrow})
\}\,. \nonumber
\end{eqnarray}
where $\rho_{R/L,\uparrow/\downarrow}$ are the charge density
operators of right/left-moving electrons with up/down-spin
projection, $V_0$ is the strength of electron-electron
interaction($V_0\sim e^2$) and $u_{1(2)}=v_{1(2)F}+V_0/2\pi\hbar$.
The Fermi velocities $v_{1F}\neq v_{2F}$ are different due to an
assumed electron dispersion asymmetry (see Fig.\ref{fig1}). We have
neglected the magnetic field-induced corrections to the Fermi
velocities and assumed that the effective electron-electron
interaction has no significant magnetic field dependence. Both the
neglected effects are of $"1/\varepsilon_F"$-order (see e.g.
Ref.~[\onlinecite{TG}]) and they are irrelevant for Zeeman
splittings $\Delta_Z\ll |\Delta| \ll \varepsilon_F$.

The Hamiltonian (\ref{llh}) is equivalent to a two-channel LL
Hamiltonian and can be diagonalized by the canonical
transformation suggested in Ref.~[\onlinecite{IVK}] (see Appendix
I). The diagonalized Hamiltonian is
\begin{equation} \label{Hd}
H_{d}=\pi\hbar \int dx \{ s_{1}(\rho_{R1}^2+\rho_{L1}^2)+
s_{2}(\rho_{R2}^2 +\rho_{L2}^2)\} ,
\end{equation}
where $s_{(1,2)}$ are the velocities of noninteracting bosonic
modes (see Appendix I).

We assume strong normal backscattering at the
S/N-boundaries (tunnel junction). In this limit the pairing Hamiltonian
contains a small factor - the amplitude of Andreev
backscattering\cite{Shel,BTK}
\begin{equation} \label{ra}
r_A^{(r,l)}\simeq D_{r,l}\exp\left[i\left(\frac{\pi}{2}+
\varphi_{r,l}\right)\right] ,
\end{equation}
where $D_{r(l)}\ll 1$ is the transparency of the barrier at the
right(left) interface, $\varphi_{r(l)}$ is the phase of the
superconducting order parameter on the right(left) bank of the
junction. The boundary Hamiltonian for our two-channel system can
be expressed in terms of the Andreev scattering amplitudes
Eq.~(\ref{ra}) up to an overall numerical factor $C$, which will
be specified later
\begin{eqnarray} \label{Hb}
H_{b}/C &=& \hbar v_{1F}\left[r_A^{\ast(l)}\Psi_{R\uparrow}(0)
\Psi_{L\downarrow}(0) - r_A^{\ast(r)}\Psi_{R\uparrow}(L)
\Psi_{L\downarrow}(L)\right] \\
&+& \hbar v_{2F}\left[r_A^{\ast(l)}\Psi_{R\downarrow}(0)
\Psi_{L\uparrow}(0) - r_A^{\ast(r)}\Psi_{R\downarrow}(L)
\Psi_{L\uparrow}(L)\right] + h.c. \nonumber
\end{eqnarray}
To second order in the Andreev scattering amplitude the phase
dependent part of the ground state energy takes the form
\begin{equation} \label{gse1}
\delta E^{(2)}(\varphi) = \sum_j \frac{|\langle j|H_b|0\rangle|^2}
{E_0-E_j} = \frac{1}{\hbar}\int_0^{\infty}d\tau\langle 0|H_b^{\dagger}(\tau)
H_b(0)|0\rangle \;,
\end{equation}
where $H_b(\tau)$ is the boundary Hamiltonian (\ref{Hb}) in the
imaginary time Heisenberg representation. After substituting
Eq.(\ref{Hb}) into Eq.(\ref{gse1}) we get the following expression
for $\delta E^{(2)}(\varphi)$ expressed in terms of electron
correlation functions
\begin{equation} \label{gse2}
\delta E^{(2)}(\varphi)=-4C\hbar\Re\left\{r_A^{\ast(l)}r_A^{(r)}
\int_0^{\infty}d\tau[v_{1F}^2\langle\Psi_{R\uparrow}(\tau,0)
\Psi_{L\downarrow}(\tau,0)\Psi_{L\downarrow}^{\dagger}(0,L)
\Psi_{R\uparrow}^{\dagger}(0,L)\rangle + v_{2F}^{2}\langle
\uparrow\Leftrightarrow\downarrow\rangle]\right\} .
\end{equation}

We will calculate the electron correlation functions in
Eq.~(\ref{gse2}) by making use of the bosonization technique.
Notice that the Zeeman splitting introduces an extra $x$-dependent
phase factor in the chiral components of the fermion fields. This
interaction can be taken into account (see e.g.
Ref.~[\onlinecite{26}]) by replacing the fermion operators in
Eq.~(\ref{gse2}) by $\Psi_{\mu,\sigma}^{(Z)}$, where
\begin{equation} \label{Z}
\Psi_{\mu,\sigma}^{(Z)} = \exp(iK_Zx)\Psi_{\mu,\sigma}\;\;\;,
\;\;\; K_Z = \frac{\Delta_Z}{4\hbar v_F}\frac{\mu\sigma-\lambda_a}
{1-\lambda_a^2}\;,
\end{equation}
Here $v_F=(v_{1F}+v_{2F})/2,\;\mu=(R,L)\equiv(1,-1)\;,
\sigma=(\uparrow,\downarrow)\equiv(1,-1)\;, \Delta_Z$ is the
Zeeman splitting, and $\lambda_a = (v_{1F}-v_{2F})/
(v_{1F}+v_{2F})$ is the parameter which characterizes the strength
of chiral symmetry breaking.

The standard bosonization formulae now read
\begin{equation} \label{bos}
\Psi_{R(L),\uparrow}(x,t)=\frac{\exp\{\pm i\sqrt{4\pi}\Phi_{R(L),\uparrow}(x,t)\}}{\sqrt{2\pi a_{1(2)}}}
\;\;\;,
\;\;\;\Psi_{R(L),\downarrow}(x,t)=\frac{\exp\{\pm i\sqrt{4\pi}\Phi_{R(L),\downarrow}(x,t)\}}{\sqrt{2\pi a_{2(1)}}}
\;,
\end{equation}
where $a_{1,2}$ are the cutoff parameters of the two-channel LL.
The chiral bosonic fields in Eq.~(\ref{bos}) for a finite length
LL are represented as follows (see e.g. Ref.~[\onlinecite{Loss}])
\begin{eqnarray}
\Phi_{R(L),\uparrow}(x,t)
&=& \frac{1}{2}\hat{\varphi}_{R(L),\uparrow} +
\hat{\Pi}_{\uparrow}\frac{x\mp v_{1(2)}t}{L_{1(2)}}+
\varphi_{R(L),\uparrow}(x,t)\;, \label{Phi1} \\
\Phi_{R(L),\downarrow}(x,t)&=&\frac{1}{2}\hat{\varphi}_{R(L),
\downarrow} + \hat{\Pi}_{\downarrow}\frac{x\mp v_{2(1)}t}{L_{2(1)}}+
\varphi_{R(L),\downarrow}(x,t)\;. \label{Phi2}
\end{eqnarray}
Here the zero mode operators
$(\hat{\varphi}_{R(L),\sigma},\hat{\Pi}_ {\sigma^{\prime}})$ obey
the commutation relations $[\hat{\varphi}_{R(L),\sigma},\hat{\Pi}_
{\sigma\prime}]=\mp i\delta_{\sigma,\sigma^{\prime}}$ and the
nontopological (harmonic) components $\varphi_{R(L),j}(x,t)$ are
\begin{equation} \label{phi}
\varphi_{R(L),j}(x,t) = \sum_{q}\sqrt{\frac{1}{2qL_j}}
\left\{e^{iq(\pm x-v_{j}t)}\hat{b}_q + h.c.\right\}\;,
\end{equation}
where $b_q(b_q^{\dagger})$ are the standard bosonic
annihilation(creation) operators.
The effective quantization lengths $L_{j}\, (j=1,2)$ depend
on the boundary conditions and will be specified in the
next section.

As is well known (see e.g. Ref.~[\onlinecite{Ners}]), the
topological excitations for an effectively infinite LL play no
role and can be omitted in Eqs.~(\ref{Phi1}) and (\ref{Phi2}).
After strightforward transformations Eq.~(\ref{gse2}) is reduced
to the following expression
\begin{equation} \label{gse3}
\delta E^{(2)}(\varphi) = 4C\hbar D\left\{v_{1F}^2
\cos(\varphi-\frac{\Delta_Z}{\Delta_{1L}})\int_0^{\infty}d\tau\Pi_{1}(\tau)+
v_{2F}^2\cos(\varphi+\frac{\Delta_Z}{\Delta_{2L}})
\int_0^{\infty}d\tau\Pi_{2}(\tau)\right\}\;\;,
\end{equation}
where $D=D_{l}D_{r}$ is the junction transparency, $\Delta_{1(2)L}=
\hbar v_{1(2)F}/L$ and
\begin{eqnarray} \label{Pi}
\Pi_{1(2)}(\tau) &=& \frac{1}{(2\pi a_{1(2)})^2}\exp\left\{2\pi\left[
\langle\langle\varphi_{\sigma}(\tau,-L)\varphi_{\sigma}\rangle\rangle
+ \langle\langle\Theta_{\rho}(\tau,-L)\Theta_{\rho}\rangle\rangle
\pm\langle\langle\Theta_{\rho}(\tau,-L)
\varphi_{\sigma}\rangle\rangle\right.\right.  \nonumber \\
&\pm&\left.\left.\langle\langle\varphi_{\sigma}(\tau,-L)
\Theta_{\rho}\rangle\rangle\right]\right\} \;.
\end{eqnarray}
Here $\varphi_{\sigma}\equiv\varphi_{\sigma}(0,0),
\Theta_{\rho}\equiv\Theta_{\rho}(0,0)$ and double brackets denote
the subtraction of the corresponding vacuum average at the points
$\tau, x=0$. The charge ($\rho$) and spin ($\sigma$) bosonic
fields in Eq.~(\ref{Pi}) are related to the chiral fields
$\varphi_{R(L),\uparrow(\downarrow)}$ introduced above by the
simple linear equations
\begin{equation} \label{le}
\varphi_{\sigma}(\Theta_{\rho}) = \frac{1}{\sqrt{2}}
\left(\varphi_{R,\uparrow}\pm\varphi_{L,\uparrow}\mp
\varphi_{R,\downarrow}-\varphi_{L,\downarrow}\right) \;
\end{equation}
(the upper sign corresponds to $\varphi_{\sigma}$ and the lower
sign denotes $\Theta_{\rho}$). With the help of the canonical
transformation Eq.~(\ref{A1}) the chiral bosonic fields in
Eq.(\ref{le}) can be expressed in terms of noninteracting
plasmonic modes $\varphi_{R/L,j}\;(j=1,2)$. For an infinitely long
LL the propagators of these fields are (see e.g.
Ref.~[\onlinecite{Ners}])
\begin{equation} \label{prop}
\langle\langle \varphi_{R/L,j}(t,x)\varphi_{R/L,k}\rangle\rangle = -\frac{\delta_{jk}}{4\pi}
\ln\frac{a_k\mp x+is_{k}t}{a_k}\,.
\end{equation}
where the velocities $s_j$ are defined in Eqs.~(\ref{A4}) and
(\ref{A5}). Finally, the expression for the Josephson current
through a side-coupled  LL wire (Fig.\ref{fig1}) takes the form
\begin{eqnarray} \label{Ji}
&&J^{(i)}(V_0,\lambda_a,\Delta_Z;\varphi)=\nonumber \\
&&\frac{ev_F}{L}D\frac{C}{2\pi^2}
\left\{
\left(\frac{a_1}{L}\right)^{2(\gamma_1-1)}\frac{v_{1F}^2}{s_1v_F}
B(1/2,\gamma_1-1/2)F(1/2,\gamma_{1s};\gamma_1;1-(s_2/s_1)^2)\sin
\left(\varphi-\frac{\Delta_Z}{\Delta_{1L}}\right)\right. \nonumber \\
&+&\left.\left(\frac{a_2}{L}\right)^{2(\gamma_2-1)}\frac{v_{2F}^2}{s_1v_F}
B(1/2,\gamma_2-1/2)F(1/2,\gamma_{2c};\gamma_2;1-(s_2/s_1)^2)\sin
\left(\varphi+\frac{\Delta_Z}{\Delta_{2L}}\right)\right\} \;,
\end{eqnarray}
where $B(x,y)=\Gamma(x)\Gamma(y)/\Gamma(x+y)$ is the beta function,
$F(\alpha,\beta;\gamma;z)$ is the hypergeometric function (see e.g.
Ref.[\onlinecite{RG}]), $v_F=(v_{1F}+v_{2F})/2\;,\;\gamma_j=\gamma_{js}
+\gamma_{jc} \,(j=1,2)$ and
\begin{equation} \label{gammas}
\gamma_{1s}=\frac{v_{2F}}{v_{1F}}\frac{\sin^2\psi}{g_2}\;,\;
\gamma_{1c}=\frac{\cos^2\psi}{g_1}\;,\;\gamma_{2s}=\gamma_{1s}(1
\leftrightarrow 2) , \gamma_{2c}=\gamma_{1c}(1\to 2) .
\end{equation}
Here $g_j=s_j/v_{jF}$ are the correlation parameters of a
two-channel LL (see Appendix I) and angle parameter $\psi$ is
defined by Eq.~(\ref{A3}).

By using the properties of the hypergeometric function it is easy
to show that for a given strength of the electron-electron
interaction the Josephson current $J^{(i)}$ satisfies the
equations
\begin{equation} \label{sym}
J^{(i)}(-\lambda_a,\Delta_Z;\varphi)=J^{(i)}(\lambda_a,-\Delta_Z;\varphi)
= -J^{(i)}(\lambda_a,\Delta_Z;-\varphi)
\end{equation}
which describe the symmetries of electric current with respect to
space and time reflections. In particular one can infer from
Eq.~(\ref{sym}) that when both the chiral symmetry breaking
($\lambda_a\neq 0$) and the Zeeman $(\Delta_Z\neq 0)$ interaction
are present the supercurrent  can persist even at $\varphi=0$.
This anomalous supercurrent exists already for noninteracting
electrons ($V_0=0$) and at first we analyze Eq.~(\ref{Ji}) in the
limit of weak e-e interaction.

For noninteracting electrons  ($V_0=0, g_1=g_2=1$) Eq.~(\ref{Ji})
is much simplified to
\begin{equation} \label{free1}
J_0^{(i)}(\varphi)=J_{c}^{(0)}\frac{1}{2}\left\{\frac{v_{1F}}{v_F}\sin
\left(\varphi-\frac{\Delta_Z}{\Delta_{1L}}\right)+\frac{v_{2F}}{v_F}\sin
\left(\varphi+\frac{\Delta_Z}{\Delta_{2L}}\right)\right\}\;
\end{equation}
where $J_{c}^{(0)}=(Dev_F/4L)(C/\pi)$ is the critical Josephson
current. We see that in the absence of magnetic interaction
($\Delta_Z=0$) the Rashba interaction in the considered geometry
of SNS junction does not affect Josephson current at all (see also
Ref.~[\onlinecite{BR}] where an analogous result was derived for a
short 2D SNS junction in the presence of Rashba spin-orbit
interactions). The interplay of the Zeeman interaction and the
dispersion asymmetry in quantum wires results in the appearance of
an anomalous (at $\varphi=0$) Josephson current $J_a^{(i)}\equiv
J_0^{(i)}(\varphi=0)$ which it is convenient to express in terms
of the asymmetry parameter $\lambda_a$ and the magnetic phase
$\chi_B=\Delta_z/\Delta_L\; (\Delta_L=\hbar v_F/L)$ as
\begin{equation} \label{ja}
 J_a^{(i)}(\lambda_a,\chi_B)=\frac{J_{c}^{(0)}}{2}\left\{
(1-\lambda_a)\sin\left(\frac{\chi_B}{1-\lambda_a}\right)-
(1+\lambda_a)\sin\left(\frac{\chi_B}{1+\lambda_a}\right)\right\} .
\end{equation}
As is evident from the above equation, the anomalous supercurrent
$ J_a$ appears only when both the dispersion asymmetry and the
Zeeman interaction are present
$\;J_a(\lambda_a=0,\Delta_Z)=J_a(\lambda_a,\Delta_Z=0)=0$. In the
limit of weak dispersion asymmetry $\lambda_a\ll 1$ (a realistic
case\cite{M1} for quantum wires formed in 2DEG) the Josephson
current as a function of Zeeman splitting demonstrates a simple
harmonic behavior with a slow periodically varying amplitude
(beats)
\begin{equation} \label{beats}
 J_a^{(i)}\simeq J_{c}^{(0)}\sin\left[\varphi+\lambda_a
\left(\frac{\Delta_Z}
{\Delta_L}-\tan\frac{\Delta_Z}{\Delta_L}\right)
\right]\cos\left(\frac{\Delta_Z}{\Delta_L}\right) .
\end{equation}

Now we analyze Eq.~(\ref{Ji}) in the limit when the Rashba
interaction is negligibly small ($\lambda_a=0$). In this case the
Josepson current through the LL wire takes the form
\begin{equation} \label{Jg}
J_g^{(i)}=J_c^{(g)}\cos\chi_B\sin\varphi\;\;\;,\;\;\;
J_c^{(g)}=R(g_c)J_{c}^{(0)} ,
\end{equation}
where the interaction-induced renormalization factor $R(g_c)\;$
(here $g_c^{-1}= \sqrt{1+2V_0/\pi\hbar v_F}$ is the LL correlation
parameter in the charge sector) is equivalent to the one evaluated
in Ref.~[\onlinecite{Fazio}]
\begin{equation} \label{Rc1}
R(g_c)=\frac{g_c}{\sqrt{\pi}}\frac{\Gamma(1/2g_c)}{\Gamma(1/2+1/2g_c))}
F\left(\frac{1}{2},\frac{1}{2};\frac{1}{2g_c}+\frac{1}{2};1-g_c^2\right)
\left(\frac{a}{L}\right)^{g_c^{-1}-1}
\end{equation}
In the limit of strong interaction $V_0/\hbar v_F\gg 1$ the renormalization
factor is small
\begin{equation} \label{Rc2}
R(g_c\ll 1)\simeq\frac{\pi}{2}\left(\frac{\hbar v_F}{V_0}\right)^{3/2}
\left(\frac{a}{L}\right)^{\sqrt{2V_0/\pi\hbar v_F}}\ll 1
\end{equation}
and the Josephson current through a S/I/LL/I/S junction is strongly
suppressed\cite{Fazio}.

When both the electron-electron interaction and the dispersion
asymmetry are strong, only one of the two terms in Eq.~(\ref{Ji})
dominates. The corresponding critical current (for definiteness we
assume that $v_{1F}\simeq v_F/2\gg v_{2F}$)
\begin{equation} \label{Jci}
J_c^{(i)}=J_{c}^{(0)}\pi\left(\frac{\hbar v_{1F}}{V_0}\right)^{3/2}
\left(\frac{a}{L}\right)^{2\sqrt{V_0/\pi\hbar v_{1F}}}
\end{equation}
is much smaller than the critical current $J_c$ in the absence
of dispersion asymmetry ($v_{1F}=v_{2F}$). It means that chiral
symmetry breaking in quantum wires enhances the interaction-induced
suppression of the Josephson current.

\section{Dispersion asymmetry and quasi-random energy spectrum of
plasmons}

In this section we evaluate the spectrum of topological
excitations and plasmonic modes in a LL wire of the length $L$
end-coupled to bulk superconductors (see Fig.\ref{fig3}). The electron
normal backscattering at the N/S interfaces is assumed to be
strong. The Josephson current can be calculated to the first order
on junction transparency using Eq.(\ref{gse2}) for the
$\varphi$-dependent part of the ground state energy. For a finite
length LL the zero modes in Eqs.~(\ref{Phi1}) and (\ref{Phi2})
contribute to the energy and after some algebra we get for $\delta
E^{(2)}(\varphi)$ an expression analogous to Eq.~(\ref{gse3})
where now $\Pi_{1(2)}(\tau)$ is replaced by the product
$\Pi_{1(2)}(\tau)Q_{1(2)}(\tau)$. The zero mode contributions
$Q_{1(2)}(\tau)$ are $(j=1,2)$
\begin{equation} \label{Qj}
Q_j(\tau) = \exp\left\{-2\pi\langle\left[\frac{L}{L_j}\left(
\hat{\Pi}_{\uparrow}- \hat{\Pi}_{\downarrow}\right) +
\frac{iv_j\tau}
{L_j}\left(\hat{\Pi}_{\uparrow}+\hat{\Pi}_{\downarrow}\right)\right]^2
\rangle\right\}\exp\left(\frac{2\pi v_j\tau}{L_j}\right).
\end{equation}
To zeroth order of perturbation theory in the barrier
transparencies the electrons are confined to the normal region.
Therefore the correlation functions in Eq.~(\ref{gse3}) have to be
calculated with the appropriated boundary conditions. The natural
boundary condition for our problem is the requirement that the
particle current through the interfaces at $x=0,L$ is zero
\begin{equation} \label{Js}
J_{\sigma}\sim
\Re\{i\Psi_{\sigma}^{\dagger}\partial_x\Psi_{\sigma}\}|_{x=0,L}=0
\;\;, \sigma=
\uparrow,\downarrow  .
\end{equation}
Here the wave function $\Psi_{\sigma}$ for the nonsymmetric
electron dispersion is represented as
\begin{equation} \label{Psi}
\Psi_{\uparrow(\downarrow)}\simeq e^{ik_{1(2)F}x}\Psi_{R1(2)}(x) +
e^{-ik_{2(1)F}x}\Psi_{L(2)1}(x) .
\end{equation}
Notice that Eqs.~(\ref{Js}) and (\ref{Psi}) determine more general
boundary conditions than $\Psi_{\sigma}(x=0,L)=0$ usually assumed
in the literature (see e.g. Ref.~[\onlinecite{NP}]). The last b.c.
is the particular case of so called "hard wall" b.c.'s
$\Psi^{(j)}(x_b)=0\;j=1,...,2N$ for a multichannel $(N)$ spin-1/2
 LL. They do not mix the channels and allows one to
reduce the multichannel problem to calculations for a single
channel situation with an additional summation of channel
dependent quantities over channel quantum numbers\cite{NP}. In our
case scattering at the boundaries changes the channel "index"
($1\leftrightarrow 2$) and the correct b.c. for "slow" fields
$\Psi_{R(L)}$ has to take this fact into account. The
decomposition Eq.~(\ref{Psi}) holds at distances much larger then
$\lambda_F$. In a general case, the wave function at the boundary
is of a more complicated and unknown form and one may not put
$\Psi_{\sigma}=0$ in order to find the relations between the two
terms in Eq.~(\ref{Psi}). In contrast, the requirement that the
particle current through the boundary is zero is robust and its
consequences hold at any distance from the boundary due to current
conservation.

For the bare electron spectrum without dispersion asymmetry
($k_{1F}= k_{2F}=k_F$) the formulated requirement is equivalent to
the following boundary conditions for the chiral (R,L) fermionic
fields (see also Ref.~[\onlinecite{FG}])
\begin{equation} \label{oe}
\Psi^{\dagger}_{R\sigma}(x)\Psi_{R\sigma}(x)|_{x=0,L} =
\Psi^{\dagger}_{L\sigma}(x)\Psi_{L\sigma}|_{x=0,L} .
\end{equation}
The boundary conditions Eq.~(\ref{oe}) correspond to a LL with
open ends\cite{FG} and result in zero eigenvalues of the
momentum-like zero-mode operator $\hat{\Pi}_{\sigma}$ and in
quantization of harmonic modes (plasmons) on a ring with
circumference $2L$ (see Ref.~[\onlinecite{FG}]). In this case the
spectrum of plasmons is equidistant and the propagators take the
form ($j,k=1,2$)
\begin{equation} \label{prop1}
\langle\langle\varphi_{R(L)j}(t,x)\varphi_{R(L)k}\rangle\rangle =
-\frac{\delta_{jk}}{4\pi}\ln\frac{1-\exp[i\pi(\pm x-s_kt+ia)/L]}
{\pi a/L}  .
\end{equation}
Here $a$ is the cutoff length and $s_{1(2)}$ are the velocities
of charge and spin excitations (for noninteracting fermions
$s_1=s_2=v_F$).

Now we generalize the quantization procedure elaborated in
Ref.~[\onlinecite{FG}] to an electron spectrum with dispersion
asymmetry. We will assume that electron normal backscattering at
the boundaries is not accompany by spin-flip processes. Therefore
each backscattering for our spectrum (Fig.\ref{fig1}) leads to the change
of the channel index ($"1"\leftrightarrow "2"$) and the
corresponding Fermi velocity.

It is worthwhile at first to consider the general case of boundary
scattering in a two-channel system of noninteracting electrons
confined to the interval [0,L]. The electron backscattering at the
boundaries is described by $2\times 2$ unitary symmetric matrix
which is convenient to parameterize as follows
\begin{equation} \label{S}
\hat{S}= e^{i\delta}\pmatrix{r&i|t|\cr i|t|&r^{\ast}} ,
\end{equation}
where $r=|r|e^{i\delta_r}$ is the intrachannel backscattering
amplitude ($1\leftrightarrow 1, 2\leftrightarrow 2$) and $t$ is
the interchannel backscattering ($1\leftrightarrow 2$) amplitude
$|r|^2 + |t|^2 =1$.  By matching the electron wave functions at
the boundaries $x=0$ and $x=L$ with the help of the S-matrix
Eq.~(\ref{S}) one easily finds the spectrum equation
\begin{equation} \label{se0}
\cos^2\left[\frac{\varepsilon L}{2}\left(\frac{1}{v_{1F}}+
\frac{1}{v_{2F}}\right)+\delta\right]=
|r|^2\cos^2\left[\frac{\varepsilon L}{2}\left(\frac{1}{v_{1F}}-
\frac{1}{v_{2F}}\right)+\delta_r\right] .
\end{equation}
For purely intrachannel reflection, $t=0$, we get from
Eq.~(\ref{se0}) two independent sets ($j=1,2$) of equidistant
levels with spacing $\Delta\varepsilon_j = \pi\hbar v_{jF}/L$. In
the opposite case of purely interchannel backscattering ($r=0$)
the spectrum is also equidistant
\begin{equation} \label{es0}
\varepsilon_n = \frac{2\pi\hbar}{L}\frac{v_{1F}v_{2F}}{v_{1F}+v_{2F}}
\left(n+\frac{1}{2}-\frac{\delta}{\pi}\right) \;\;\;, n=1,2,...
\end{equation}
In a general case the spectral equation (\ref{se0}) yields a set
of quasi-random energy levels.

The bozonization technique is compatible only with the two
considered limiting cases: $|r|=1$ (this was demonstrated in
Ref.~[\onlinecite {FG}]), and $r=0$ as we will show now. Let us
start at first with the case on noninteracting fermions. The
boundary condition Eq.~(\ref{Js}) for $v_{1F}\neq v_{2F}$ results
in the equations
\begin{equation} \label{bc1}
v_{1F}\Psi^{\dagger}_{R1}(x)\Psi_{R1}(x)|_{x=0,L}=
v_{2F}\Psi^{\dagger}_{L2}(x)\Psi_{L2}(x)|_{x=0,L}  ,
\end{equation}
\begin{equation} \label{bc2}
\Re\left[\Psi^{\dagger}_{L2}(x)\Psi_{R1}(x)e^{i(k_{1F}+k_{2F})x}
\right]_{x=0,L}=0 .
\end{equation}
These equations are satisfied if
\begin{equation} \label{sol1}
\frac{a_1}{a_2}=\frac{L_1}{L_2}=\frac{v_{1F}}{v_{2F}}\;\;,\;\;
\frac{1}{L_1}+\frac{1}{L_2}=\frac{1}{L}
\end{equation}
\begin{equation} \label{sol2}
\varepsilon_n^F = \frac{2\pi}{L}\frac{v_{1F}v_{2F}}{v_{1F}+v_{2F}}n
\;\;,\;\;n=1,2,...
\end{equation}
and
\begin{equation} \label{sol3}
[\Phi_{L\sigma}(x,t)+\Phi_{R\sigma}(x,t)]|_{x=0,L}= \frac{\sqrt{\pi}}
{2}n_{\sigma}\;\;,\;\;\sigma=\uparrow,\downarrow  ,
\end{equation}
where $n_{\uparrow}$ and $n_{\downarrow}$ are integers.
Eq.~(\ref{sol3}) in its turn is satisfied for topological sector
with quantum numbers
$\;(\hat{\varphi}_{R\sigma}+\hat{\varphi}_{L\sigma})/\sqrt{\pi}=
n_{\sigma},\;\hat{\Pi}_{\sigma}=0$ and the harmonic modes
$\varphi_{R(L)\sigma}(x,t)$ which obey the relations
\begin{equation} \label{hbc}
\varphi_{R\sigma}(x,t)|_{x=0,L}=-\varphi_{L\sigma}(x,t)|_{x=0,L} .
\end{equation}
>From Eqs.~(\ref{phi}), (\ref{sol1}) and (\ref{hbc}) one easily
gets the plasmon spectrum
\begin{equation} \label{en}
\varepsilon_n=\frac{2\pi}{L}\frac{s_{1}s_{2}}{s_1+s_2}n
\end{equation}
($s_{1,2}$ are the plasmon velocities, which coincide with the
Fermi velocities for noninteracting fermions) and the desired
correlation functions ($j,k=1,2$)
\begin{equation} \label{prop2}
\langle\langle\varphi_{R(L)j}(x,t)\varphi_{R(L)k}\rangle\rangle=
-\frac{\delta_{jk}}{4\pi}\ln\frac{1-\exp[i2\pi(\pm x-s_kt+ia_k)/L_k]}
{2\pi a_k/L_k} ,
\end{equation}
where the effective quantization lengths $L_j$ according to
Eq.~(\ref{sol1}) are
\begin{equation} \label{Lj}
L_{1(2)}=\frac{v_{1F}+v_{2F}}{v_{2(1)F}}L
\end{equation}
In the limit $v_{1F}=v_{2F}$ Eqs.~(\ref{en})-(\ref{Lj}) reproduce
the plasmon spectrum and the correlation functions of a single
channel LL with open ends\cite{FG}.

Now we are ready to consider the effects of interaction. For a
single-mode LL (or for a multichannel LL, provided the
backscattering is allowed only to its own channel) the boundary
condition Eq.~(\ref{hbc}) for harmonic modes holds also for
interacting fermions as one can check using a Bogoliubov-like
transformation which diagonalizes the LL Hamiltonian. Hence in the
presence of interaction one can still use the same correlation
functions as for noninteracting fermions with the only difference
that the velocities are renormalized by interaction.

This is not the case for our problem. With the help of exact
transformations (see Eq.~(\ref{A1})) which diagonalize the
2-channel LL Hamiltonian\cite{IVK} one can show that if the chiral
bosonic fields satisfy Eq.~(\ref{hbc}), the diagonalized ones
$\tilde{\varphi}_{R(L)j}$ are connected at the boundaries by the
effective "scattering matrix" $\hat{S}^e$
\begin{equation} \label{Se}
\tilde{\varphi}_{Rj}(x=0,L)=\sum_{k=1}^{k=2}S^e_{jk}\tilde{\varphi}_
{Lk}(x=0,L)\;\;\;,\;\;\;\hat{S}^e=\frac{1}{B}\pmatrix{-A&1\cr 1&A} ,
\end{equation}
where the coefficients $A,B$ are
\begin{eqnarray} \label{A}
A=-&[\cos2\psi-\sinh(\vartheta_1-\vartheta_3)\sin2\psi]^{-1}
[\sinh(\vartheta_1-\vartheta_2)\cosh(\vartheta_1-\vartheta_3) \\ \nonumber
+&\cos2\psi\cosh(\vartheta_1-\vartheta_2)\sinh(\vartheta_1-\vartheta_3)
+\sin2\psi\cosh(\vartheta_1-\vartheta_2)] ,
\end{eqnarray}
\begin{eqnarray} \label{B}
B=-&[\cos2\psi-\sinh(\vartheta_1-\vartheta_3)\sin2\psi]^{-1}
[\cosh(\vartheta_1-\vartheta_2)\cosh(\vartheta_1-\vartheta_3) \\ \nonumber
+&\cos2\psi\sinh(\vartheta_1-\vartheta_2)\sinh(\vartheta_1-\vartheta_3)
+\sin2\psi\sinh(\vartheta_1-\vartheta_2)]
\end{eqnarray}
and the "rotation angles" $\vartheta_l\;(l=1,...4)$ and $\psi$ are
defined in Appendix I (see Eqs.~(\ref{A2}) and (\ref{A3})). One
can check after some algebra that the coefficients $A$ and $B$
satisfy the simple relation $B^2-A^2=1$, which makes the S-matrix
in Eq.~(\ref{Se}) unitary. This observation allows us to use ~the
scattering matrix formalism when evaluating the energy spectrum of
plasmons. Notice that in the parametrization Eq.~(\ref{S}) we have
$r=iA/B,\;|t|=1/B,\;\delta=\pi/2$.

For monochromatic bosonic fields with amplitudes $b_{R(L)j}$ the
scattering at the boundaries $x=0$ and $x=L$ are determined by
the equations
\begin{equation} \label{bj}
x=0:\;\;b_{Rj}=\sum_{k=1}^2 S^e_{jk}b_{Lk}\;\;\;,\;\;\;
x=L:\;\;e^{-i\alpha_j}b_{Lj}=\sum_{k=1}^2 S^e_{jk}e^{i\alpha_k}b_{Rk} ,
\end{equation}
where the phases $\alpha_j=\varepsilon L/s_j$ and $s_j$ are the plasmon
velocities (see Eqs.(\ref{A4}),(\ref{A5})).
>From the above set of linear equations
one easily finds the spectrum equation for plasmons
\begin{equation} \label{ps}
\sin^2\left[\frac{\varepsilon
L}{2}\left(\frac{1}{s_1}+\frac{1}{s_2}\right)\right]=
R\sin^2\left[\frac{\varepsilon L}{2}\left(\frac{1}{s_1}-\frac{1}{s_2}
\right)\right] ,
\end{equation}
where $R\equiv (A/B)^2\leq 1$ is the effective backscattering
coefficient for plasmons. It depends both on the dimensionless
 interaction strength $\kappa=V_0/\pi\hbar(v_{1F}+v_{2F})$ and
on the dispersion asymmtry parameter $\lambda_a$. Notice that the
spectral equation (\ref{ps}) is the special case of
Eq.~(\ref{se0}) for $\delta=\delta_r =\pi/2$.

The derived spectral equation has simple exact analytic solutions
in two limiting cases: (i) noninteracting fermions and, (ii) when
dispersion asymmetry is absent, $v_{1F}=v_{2F}=v_F$. For
noninteracting particles ($V_0=0$ in Eq.~(\ref{llh})) the
"rotation angles" are $\psi=0, \vartheta_1=\vartheta_4=0$ (see
Appendix I) and the velocities $s_1=v_{1F}, s_2=v_{2F}$. Then
$A=0, B=-1$ and the effective backscattering coefficient $R=0$.
Eq.~(\ref{ps}) in this limit reproduces the equidistant spectrum
of "noninteracting" plasmons, Eq.~(\ref{es0}). For interacting
fermions in the absence of dispersion asymmetry the "rotation
angles" are $\vartheta_1=\vartheta_3, \vartheta_2=\vartheta_4,
cos2\psi=0$. In this limit $s_1=s, s_2=v_F$ and $R=1$. The
corresponding plasmon energies $\varepsilon_n^{(1)}=\pi sn/L,\;
\varepsilon_n^{(2)}=\pi v_Fn/L , (n=1,2,...)$ represent the
standard excitations in the  charge and spin sector of a finite
length ($L$) Luttinger liquid with open ends\cite{FG}.

For a general case Eq.~(\ref{ps}) has to be solved numerically and
the plasmon spectrum represents a set of quasi-random energy
levels. The plasmonic energies can not be separated into two
independent sets of levels - one for charge density excitations,
another for spin density excitations. It means that the considered
boundary conditions strongly mix the charge and spin excitations
and the phenomena of charge-spin separation, well known in a LL,
generally speaking, can disappear when both spin-orbit interaction
and finite size effects are present.

\section{ Josephson current through a finite-length LL wire with
dispersion asymmetry}

It is clear from a physical point of view that the effects of a
dispersion asymmetry in the bare electron spectrum have to be most
significant in the quantum dynamics of noninteracting electrons.
In this case the mismatch in Fermi velocities when an electron is
backscattered at the boundaries leads to an intricate interference
pattern. The more strongly particles interact, the less important
are the effects of dispersion asymmetry. For instance, in the
limiting case of a 1D Wigner crystal (strong repulsive long-range
interactions) it is hard to imagine any interference produced by
the quantum dynamics of plasmons in two Wigner crystals pinned by
structural imperfections at the boundaries. So in our problem it
is reasonable to expect the restoration of the regular plasmon
spectrum and the spin-charge separation in the limit of strong
interaction.

For strongly interacting electrons
$\kappa=V_0/\pi\hbar(v_{1F}+v_{2F}) \gg 1$ and $v_{1F}\sim v_{2F}$
(i.e. for week or moderate dispersion asymmetry) the coefficient
$R$ (intra-channel backscattering probability) in Eq.~(\ref{ps})
takes the form
\begin{equation} \label{R1}
R\simeq 1 -\frac{1}{\kappa^{3/2}}\frac{\lambda_a^2(4-3\lambda_a^2)}
{2\sqrt{1-\lambda_a^2}} .
\end{equation}
We see that when $\kappa\gg 1$ the backscattering is always an
intrachannel process ($R\simeq 1$ with a high accuracy) and the
spin-charge separation and the equidistant character of plasmon
spectra are indeed restored. This observation (see also
Ref.~[\onlinecite{IVK}]) justifies for strongly interacting
multichannel ($j=1,...N$)
 LL the boundary conditions
($\Psi^{(j)}_{\uparrow,\downarrow}(x=0,L)=0$) usually postulated
in the literature (see e.g. Ref.~[\onlinecite{NP}]) for arbitrary
interaction.

It is straightforward to evaluate the Josephson current using the
exact plasmon spectrum for $R=1$ (i.e. when $\lambda_a=0$) and the
propagators Eq.~(\ref{prop1}). The result for zero Zeeman
splitting ($\Delta_Z=0$) is
\begin{equation} \label{jf}
J^{(f)}(g_c;\varphi)=J_c^{(0)}R_f(g_c)\sin\varphi  ,
\end{equation}
where $J_c^{(0)}=(Dev_F/4L)(C/\pi)$ is the critical current
through a 1D SNS junction and the interaction-induced
renormalization factor $R_f(g_c)$ is
\begin{equation} \label{Rf}
R_f(g_c)=\frac{2g_c^2}{2-g_c^2}F(2g_c^{-1},2g_c^{-1}-g_c;2g_c^{-1}+
1;-1)\left(\frac{\pi a}{L}\right)^{2(g_c^{-1}-1)} .
\end{equation}
Here $F(\alpha,\beta;\gamma;z)$ is the hypergeometric function and
$g_c$ is the LL correlation parameter (see Eq.~(\ref{Jg})). For
noninteracting electrons $R_f(g_c=1)=1$ and our formula has to
reproduce the known expression for the Josepson current through a
1D SNS junction (see e.g. Ref.~[\onlinecite{Shum}]). From this
comparison one finds $C=\pi$.

As we have already mentioned in this section, $R\rightarrow 1$ in
the limit of strong interaction $\kappa\gg 1$. This observation
allows us to evaluate the correlation functions and the Josephson
current for strongly interacting electrons with dispersion
asymmetry. The Josephson current is described by Eqs.~(\ref{jf})
and (\ref{Rf}) after the replacement $g_c\rightarrow
\kappa^{-1/2}$ and in the limit $\kappa\gg 1$. The renormalization
factor now takes the form
\begin{equation} \label{Rf1}
R_f(\kappa\gg 1)\simeq \frac{1}{\kappa}\left(\frac{\pi a}{L}
\right)^{2\sqrt{\kappa}}\ll 1 .
\end{equation}
The formulae (\ref{jf}) and (\ref{Rf1}) show that in the
considered limit the Josephson current dos not depend on the
parameter $\lambda_a$ of dispersion asymmetry. By comparing
Eq.~(\ref{Rf1}) and Eq.~(\ref{Jci}) we see that the interaction
suppresses supercurrent more strongly in a long end-coupled
quantum wire then in a side-coupled one, which is in agreement
with intuition.

Dispersion asymmetry affects the supercurrent of weakly
interacting electrons. The influence, however, numerically is not
strong even for the most favourable case of noninteracting
particles. With the help of the correlation functions
(\ref{prop2}) it is easy to calculate the Josephson current of
noninteracting electrons with dispersion asymmetry
\begin{equation}
J^{(f)}(\lambda_a,\Delta_Z;\varphi) = J_c^{(0)}R(\lambda_a)
\cos\left[\frac{\Delta_Z}{2}\left(\frac{1}{\Delta_{L1}}+
\frac{1}{\Delta_{L2}}\right)\right] .
\end{equation}
Here $J_c^{(0)}$ is the critical current in the absence of
dispersion asymmetry (see Eq.~(\ref{jf})) and $R(\lambda_a)$ is
the renormalization factor induced by the asymmetry of electron
dispersion
\begin{equation} \label{Ra}
R(\lambda_a)=\frac{\pi\lambda_a(1-\lambda_a^2)}{\sin(\pi\lambda_a)}
\simeq\left\{
\begin{array}{ll}
1+(\pi^2/6-1)\lambda_a^2 \;\;\;&\lambda_a\ll 1 \\
2                              &\lambda_a\rightarrow 1
\end{array}
\right.
\end{equation}
We see from Eq.~(\ref{Ra}) that the dispersion asymmetry always
slightly enhances the critical current. The analysis of the
Josephson current in a 1D SQWS junction in the presence of
dispersion asymmetry was performed in Ref.~[\onlinecite{KGSJ}]
using the Andreev level approach. It was shown that the observed
enhancement of the Josephson current is due to less perfect
cancellations (different Fermi velocities) of the partial
supercurrents carried by adjacent Andreev levels.

\section{The Rashba effect, chiral electrons in 1D quantum wires
and the Josephson current in SLLS junction}

Now we consider the limit of strong Rashba interaction. In this
case the electrons in a quasi-1D quantum wire behave like truly
chiral particles, that is the spin polarization of an electron
irrespective of its  subband index is determined by the direction
of electron motion along the wire --- right-moving and left-moving
electrons have opposite spin projections\cite{GZ}. We will assume
for definiteness (it depends on the sign of Rashba coupling) that
"R"-electrons are "down"-polarized and "L"-electrons are
"up"-polarized (see Fig.\ref{fig3}, dashed lines for spin polarizations).
We have already seen in section II, that the left/right symmetry
breaking in the presence of the Zeeman interaction results in the
appearance of an anomalous Josephson current. Physically it means
that when the spin projection is correlated with the direction of
motion (left, right), the magnetic field, via the Zeeman
interaction, induces partial Josephson currents (for each subband
"1" and "2") even if the superconducting phase difference in the
SNS junction is zero.
 For the spin alignments assumed in Refs.[\onlinecite{M1,M2}]
the subbands contribute to the Josephson current with opposite
signs and therefore the anomalous supercurrent vanishes for
symmetrical spectrum $v_{1F}=v_{2F}$. As we see, the electron dispersion
asymmetry is indispensible property to get anomalous Josepson
current for a weak Rashba interaction. In the limit of strong
s-o interaction when all right(left) moving particles have
parallel spins, the contributions of subbands have the same
sign and the existence of electron dispersion asymmetry
ceases to be crucial in appearence of anomalous (at $\varphi
=0$) Josepson current.

What is more important are the spin-flip processes which may take
place in the transition regions between the 2D or 3D electron
reservoirs (superconducting leads in our case) and the 1D quantum
wire with a pronounced Rashba effect. The electrons in the
reservoirs have two possible spin states, while deep inside the
wire, where the s-o interaction is strong, the electron spins have
to be aligned according to the above discussed prescription. So
particles with the "wrong" spin projection should turn their spins
toward the "right" direction.

One can imagine two different types of transition regions. In the
case when s-o interaction is changed abruptly at the lead/wire
interfaces, the spin-flips induced by the Rashba interaction will
be accompanied by normal electron backscattering. Such
non-adiabatic contacts were studied in Ref.~[\onlinecite{GZ}] when
evaluating normal electron transport through a 1D quantum wire
with strong Rashba interaction attached to leads where the s-o
interaction was assumed to be negligibly small. In this model the
transparency of the junction strongly depends on the spin-orbit
coupling and normal electron transport is suppressed even for
perfect contacts.

Here we will assume that the Rashba interaction in the QW is
switched on smoothly and that therefore the backscattering of the
electron at the boundary and the rotation of its spin induced by
the Rashba interaction the in quasi-1D quantum wire are spatially
separated. For instance, the left-moving electron (spin-"up") at
first is normally backscattered at the left interface keeping the
spin-"up" and then after travelling some length $\lambda_F\ll l\ll
L$  its spin becomes "down"-polarized. In this model, the Rashba
interaction does not lead to additional electron backscattering at
the interfaces and does not suppress the supercurrent.

It is also convenient for calculations to assume that the
electron-electron interaction in the 1D QW is strong. As was shown
in section IV, the interchannel electron backscattering at the
I/LL interfaces is suppressed in the limit of strong repulsive
interaction and one can use a simple quantization procedure
proposed in Ref.~[\onlinecite{FG}] to evaluate the plasmon
spectrum and the correlation functions. After straightforward
calculations the desired expression for the Josephson current
takes the form
\begin{equation} \label{JR}
J^{(R)}(\varphi)\simeq J_c^{(0)}R_{f}\sin\left[\varphi+
\frac{\Delta_Z}{2}\left(\frac{1}{\Delta_{L1}}+
\frac{1}{\Delta_{L2}}\right)\right]
\cos\left[\frac{\Delta_Z}{2}\left(\frac{1}{\Delta_{L1}}-
\frac{1}{\Delta_{L2}}\right)\right] ,
\end{equation}
where the interaction-induced renormalization coefficient $R_{f}$
is determined by Eq.~(\ref{Rf1}). As was already evident from
physical considerations, the anomalous supercurrent
$J^{(R)}(\varphi=0)$ in the limit of strong Rashba interaction is
induced by a magnetic field ($\Delta_Z\neq 0$) even in the absence
of electron dispersion asymmetry. We see from Eq.~(\ref{JR}) that
the dependence of the supercurrent on magnetic field is absolutely
different for chiral and normal electrons. In particular the
critical current for symmetric electron spectrum ($v_{1F}=v_{2F}$)
in the case of chiral electrons does not at all depend on the
Zeeman splitting, while in the ordinary situation one gets a
periodic dependence.

\section{Conclusion}

The problem we have studied allows one to consider the interplay
of proximity-induced superconductivity, the Rashba, Zeeman and
Coulomb interactions on the transport properties of quasi-1D
quantum wires. We have shown that the Rashba and Zeeman effects
strongly influence the supercurrent. In particular, the Rashba
effect in quantum wires results in a strong correlation between
electron spin polarization and the direction of electron motion.
In other words a strong Rashba interaction creates chiral
particles in the 1D electron system. The influence of a magnetic
field via the Zeeman interaction on chiral particles leads to the
appearance of a net electric current in the wire. When the leads
that the quantum wire is attached to are superconducting this
current persists even at zero phase difference across the
junction. The effect exists already for noninteracting particles
and it is strongly sensitive to the electron dispersion asymmetry
for weak Rashba coupling and ceases to depend on the asymmetry
parameter in the regime of strong Rashba interactions.

It is well known\cite{Maslov,Zag} that the Josephson current in a
perfectly transmitting junction (i.e. without normal electron
backscattering) is not influenced by the Coulomb interaction. In
contrast, any potential barrier inside the normal region which
induces electron backscattering is renormalized (upwards) by the
repulsive interaction (the Kane-Fisher effect\cite{KF}) and
therefore strongly suppresses the supercurrent through a (poorly
transmitting) SILLIS junction \cite{Fazio,Maslov,Zag,NP}. We have
shown that the electron dispersion asymmetry, which is induced by
the Rashba interaction in quasi-1D quntum wires\cite{M1,M2} is
significant for the superconducting properties of a LL junction
only for weak or moderate Coulomb interaction. In this case the
interplay of interaction and dispersion asymmetry leads to an
intricate interference pattern in the plasmon quantum dynamics of
a finite length two-channel LL and makes the plasmon spectrum
quasi-random. Strong Coulomb interactions suppress this kind of
quantum behavior and restores a regular (equidistant) plasmon
spectrum. The tendency of strong Coulomb interactions to suppress
quantum interference can be traced in different 1D electronic
systems, for instance, in a LL double barrier (absence of resonant
tunnelling for a strong repulsive interaction\cite{res}) or in
mesoscopic coupled rings (ordering effect of Coulomb interaction
on persistent current oscillations\cite{Can}).
\begin{center}
{\bf Acknowledgments}
\end{center}

This research is supported by the Royal Swedish Academy of
Sciences (KVA) and by the Swedish Research Council (LIG,RIS).
The authors thanks E.~Bezuglyi, Yu. Galperin, L.~Gorelik 
and V.~Shumeiko for
numerous fruitful discussions. IVK and AK acknowledge the
hospitality of the Department of Applied Physics at Chalmers
University of Technology and G\"{o}teborg University, Sweden. AK
also acknowledges the hospitality of the Theoretische Physik III
Insitut, Ruhr-Universit\"{a}t Bochum, Germany. IVK gratefully
acknowledges discussions with B.~Altshuler, L. Glazman, V.~Kravtsov and
A.~Nersesyan, and the hospitality and the financial support from
the Abdus Salam ICTP (Trieste, Italy),  where this work was
completed.

\newpage
\appendix{{\bf Appendix I}}

The canonical pseudoorthogonal transformations, which diagonalize
the Luttinger liquid Hamiltonian (\ref{llh}) are\cite{IVK}
\begin{equation} \label{A1}
\pmatrix{\rho_{R\uparrow}\cr\rho_{L\downarrow}\cr\rho_{R\downarrow}\cr
\rho_{L\uparrow}}=
\pmatrix{
\cosh\vartheta_1\cos\psi & \sinh\vartheta_1\cos\psi & -\cosh\vartheta_2
\sin\psi & -\sinh\vartheta_2\sin\psi \cr
\sinh\vartheta_1\cos\psi & \cosh\vartheta_1\cos\psi & -\sinh\vartheta_2
\sin\psi & -\cosh\vartheta_2\sin\psi \cr
\cosh\vartheta_3\sin\psi & \sinh\vartheta_3\sin\psi & \cosh\vartheta_4
\cos\psi & \sinh\vartheta_4\cos\psi \cr
\sinh\vartheta_3\sin\psi & \cosh\vartheta_3\sin\psi & \sinh\vartheta_4
\cos\psi & \cosh\vartheta_4\cos\psi}
\pmatrix{\rho_{R1}\cr\rho_{L1}\cr\rho_{R2}\cr\rho_{L2}} ,
\end{equation}
where the "rotation angles" $\vartheta_j$ and $\psi $ are
expressed in terms of the Fermi velocities $v_{1F}, v_{2F}$ and
the interaction strength $V_0$ by the following equations
\begin{equation} \label{A2}
\vartheta_1=\frac{1}{2}\ln g_1 ,\;\;\;\vartheta_2=\frac{1}{2}\ln
\left(\frac{v_{1F}}{v_{2F}}g_2\right) ,\;\;\;\vartheta_3=
\frac{1}{2}\ln\left(\frac{v_{2F}}{v_{1F}}g_1\right) ,\;\;\;
\vartheta_4=\frac{1}{2}\ln g_2 ,
\end{equation}
\begin{equation} \label{A3}
\tan 2\psi=\frac{2V_0\sqrt{v_{1F}v_{2F}}}{(v_{1F}-v_{2F})
[V_0+\pi\hbar(v_{1F}+v_{2F})]} .
\end{equation}
Here $g_j=v_{jF}/s_j \;(j=1,2)$ are the correlation parameters of
a 2-channel LL and the plasmon velocities are
\begin{equation} \label{A4}
s_1=v_{1F}\left\{\cos^2\psi+\left(\frac{v_{2F}}{v_{1F}}\right)^2
\sin^2\psi+\frac{V_0}{\pi\hbar v_{1F}}\left(\cos\psi+
\sqrt{\frac{v_{2F}}{v_{1F}}}\sin\psi\right)^2\right\}^{1/2} ,
\end{equation}
\begin{equation} \label{A5}
s_2=s_1(\psi\leftrightarrow -\psi, v_{1F}\leftrightarrow v_{2F}).
\end{equation}
For noninteracting electrons, $V_0=0$, the correlation parameters
are $g_1=g_2=1$ and, according to Eqs.~(\ref{A2}) and (\ref{A3})
$\vartheta_1= \vartheta_4=0,\;\psi=0$. In the limit
$v_{1F}=v_{2F}=v_F$ Eqs.~(\ref{A2})- (\ref{A5}) reproduce the
well-known expressions for the correlation parameters of a
spin-1/2 LL
\begin{equation}
\vartheta_1=\vartheta_3=\frac{1}{2}\ln g_c,\;\;\;\vartheta_2=\vartheta_4
=0,\;\;\;g_c=\left(1+\frac{2V_0}{\pi\hbar v_F}\right)^{-1/2} .
\end{equation}

\end{document}